\documentclass[preprint,superscriptaddress]{revtex4-1}
\usepackage{amsmath}
\usepackage{graphicx}

\begin{document}

\title{Rational proofs for quantum computing}
\author{Tomoyuki Morimae}
\email{tomoyuki.morimae@yukawa.kyoto-u.ac.jp}
\affiliation{Yukawa Institute for Theoretical Physics,
Kyoto University, Kitashirakawa Oiwakecho, Sakyoku, Kyoto
606-8502, Japan}
\affiliation{JST, PRESTO, 4-1-8 Honcho, Kawaguchi, Saitama,
332-0012, Japan}
\author{Harumichi Nishimura}
\email{hnishimura@is.nagoya-u.ac.jp}
\affiliation{Graduate School of Informatics,
Nagoya University, Furocho, Chikusaku, Nagoya, Aichi,
464-8601, Japan}

\date{\today}
\begin{abstract}
It is an open problem whether a classical client can
delegate quantum computing to an efficient
remote quantum server in such a way
that the correctness of quantum computing is somehow guaranteed.
Several protocols for verifiable delegated 
quantum computing have been proposed,
but the client is not completely free from any quantum technology:
the client has to generate or measure single-qubit states.
In this paper, we show that the client can be completely classical
if the server is rational (i.e., economically motivated),
following the ``rational proofs" framework of Azar and Micali.
More precisely, we consider the following protocol.
The server first sends the client a message allegedly
equal to the solution of the problem that the client
wants to solve. The client then gives the server a 
monetary reward whose amount is 
calculated in classical probabilistic polynomial-time by using
the server's message as an input.
The reward function is constructed in such a way that
the expectation value of the reward
(the expectation over the client's probabilistic computing)
is maximum when the server's message is the correct solution
to the problem.
The rational server who wants to maximize his/her profit
therefore has to send the correct solution to the client.

\end{abstract}
\maketitle

\section{Introduction}
\begin{figure}[htbp]
\begin{center}
\includegraphics[width=0.3\textwidth]{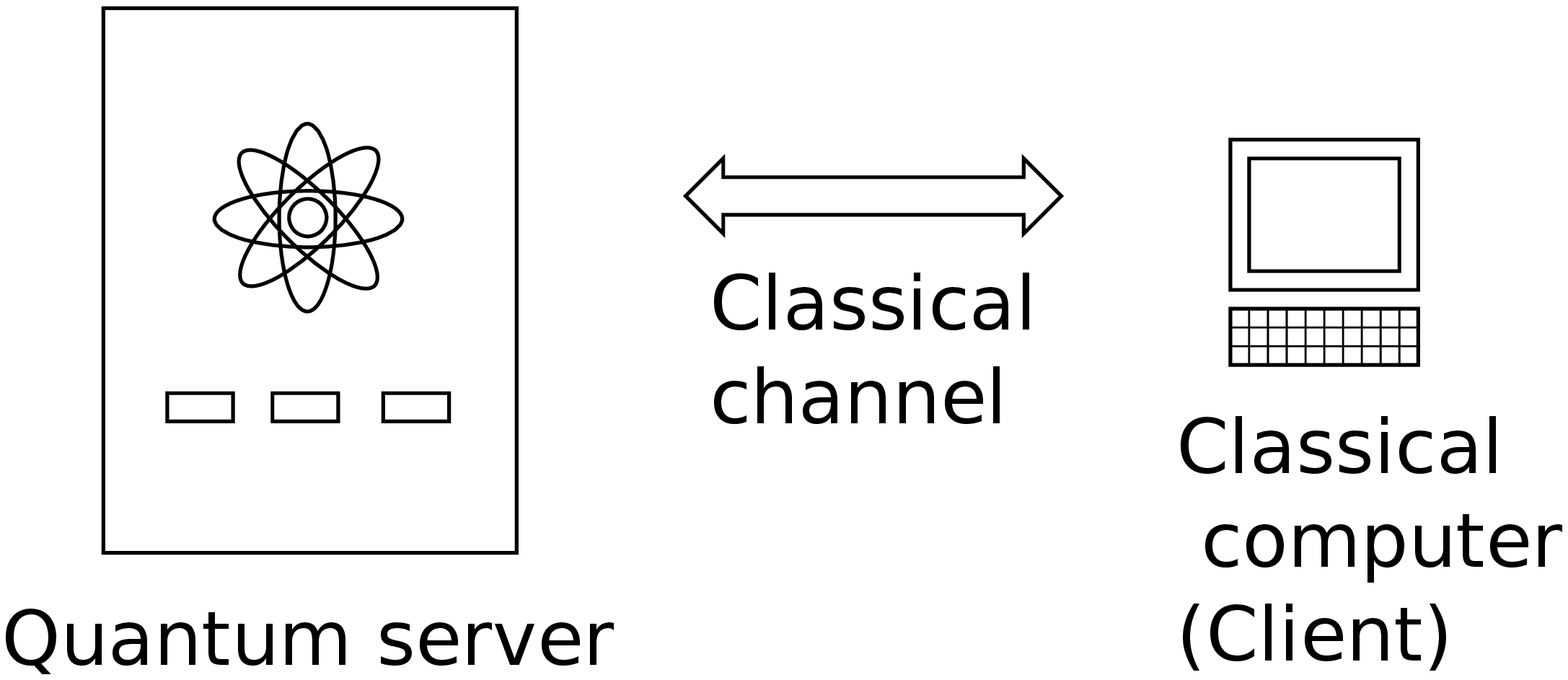}
\end{center}
\caption{
The classical verification of quantum computing.
} 
\label{setup}
\end{figure}

One of the most important open problems in
quantum physics and
quantum computing is 
the possibility of classically verifying
quantum computing~\cite{Gottesman,AharonovVazirani,Andru_review}.
As is shown in Fig.~\ref{setup},
the client, a classical computer, is connected
to a remote quantum server via a classical channel.
The server does quantum computing for the client,
and sends the result to the client.
The client, who does not trust the server, needs some
guarantee that the result is correct.
How can the correctness of server's
quantum computing be guaranteed?
There is an ironical dilemma here: 
quantum computing is useful because it cannot be classically
efficiently simulated, but exactly because of the fact,
it is impossible for the client
to verify the correctness of server's quantum computing via
the direct classical simulation.

So far, five different types of approaches
have been taken to the open problem.
First, if the client is allowed to be ``slightly quantum",
verifiable delegated quantum computing is possible.
For example, verification protocols of Refs.~\cite{MNS,posthoc} 
and verifiable blind quantum computing 
protocols~\cite{plugin,FK,Aharonov,HM,Broadbent,Andru_review,Barz,Chiara,
Andru,TakeuchiMorimae,MorimaeTakeuchiHayashi,Imoto}
assume some minimum quantum technologies for
the client, such as small quantum memories, 
single-qubit state generations, or
single-qubit measurements.

Second, if multiple entangling quantum servers
who are not communicating with each other are 
allowed, a completely classical client can verify
the correctness of servers' quantum computing~\cite{MattMBQC,RUV,Ji}.

Third, several specific problems solvable with quantum computing
have been shown to be classically verifiable.
For example, 
Simon's problem~\cite{Simon} and
factoring~\cite{Shor} are trivially classically verifiable.
Furthermore, the recursive Fourier sampling~\cite{BV}
has a poly-round-message-exchange verification protocol
between a single quantum
server and a completely classical client~\cite{MattFourier}.
Certain problems regarding the output probability
distributions of quantum circuits
in the second level of the Fourier hierarchy~\cite{FH} 
have single-message
verification protocols~\cite{Tommaso,FH2}.
Calculating the order of solvable groups has
two or three-message verification protocols~\cite{Francois}.

Fourth, 
it is known that quantum computing is verifiable by using
a technique so called the sum check protocol.
However, if we use the sum check protocol,
the server needs much stronger computational power than
usual quantum computing. (More precisely, the servers needs
to be $\#$P~\cite{complexityzoo}, which is believed to be much 
stronger than NP.)
In Ref.~\cite{AharonovGreen},
authors constructed a ``quantum version" of the sum check protocol so that
the computational power of the server becomes weaker (but still
stronger than usual quantum computing).

Finally, a recent innovative work has shown that
a classical verification of quantum computing is indeed possible
with the assumption that
the learning with errors problem is hard for quantum computing~\cite{Mahadev}.

\begin{figure}[htbp]
\begin{center}
\includegraphics[width=0.3\textwidth]{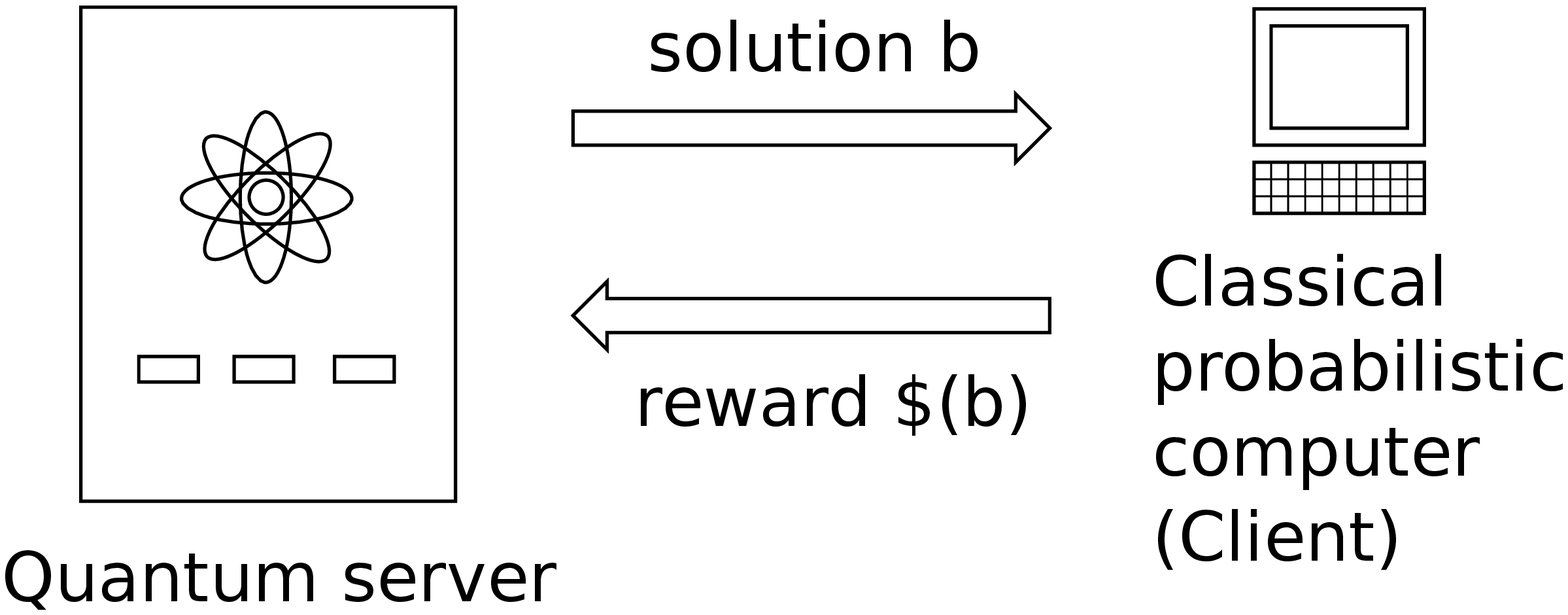}
\end{center}
\caption{
The quantum rational proof system.
} 
\label{setup2}
\end{figure}

In this paper, we take a new approach different from these previous
works. We consider a delegated quantum computing with
a rational server.
As is shown in Fig.~\ref{setup2},
the server first sends the client a message $b$ allegedly
equal to the solution of the problem that the client
wants to solve. The client then does a classical 
probabilistic polynomial-time computing to calculate
a reward $\$(b)$,
and pays $\$(b)$ to the server.
The reward function $\$$ is constructed in such a way that
the expectation value of $\$$
(over the client's probabilistic computing)
is maximum when $b$ is the correct solution.
Therefore, the rational server who wants to maximize his/her 
profit has to choose $b$ as the correct solution.

We propose two protocols.
The first protocol 
is for decision problems solvable with polynomial-time
quantum computing (i.e., BQP).
(Actually, the same construction works also for
other classes in PP.)
The second protocol 
is for estimating output probability distributions of
quantum circuits.
Finally, some discussions are given.

The idea of the rational server
was first introduced by Azar and Micali
in the context of (classical) interactive proof systems~\cite{AzarMicali}, 
which is called ``rational proof systems".
(In Sec.~\ref{App:AM}, we provide a brief summary of their 
results. Although understanding their results is not necessary to
understand our results, we provide them because they
are insightful, and therefore should be useful for
readers.)
Among several results, Azar and Micali constructed rational proof systems
for $\#$P and PP.
We bring the idea of the rational proof systems to
the verification of quantum computing.
To our knowledge, it is the first time that the 
concept of the rational proof systems is applied to quantum information.
We believe that the rational proof systems will also be useful
in many other areas of quantum information than the
verification of quantum computing.

\section{Results}
\subsection{First protocol}
\label{Sec:1stprotocol}
In this subsection,
we propose our first protocol for BQP.
Let us assume that
the client wants to solve a decision problem $L$ in BQP.
The client asks the server to solve the problem,
and the server sends a single bit $b\in\{0,1\}$
to the client. If the server is honest,
the server sends the client $b=1$ when the answer is yes 
$(x\in L)$,
and $b=0$ when 
the answer is no $(x\notin L)$.

Since BQP is in PP, there exists a classical probabilistic
polynomial-time algorithm $A$ that outputs $a\in\{0,1\}$ such that
\begin{itemize}
\item
If $x\in L$ then ${\rm Pr}[a=1]>\frac{1}{2}$.
\item
If $x\notin L$ then ${\rm Pr}[a=1]<\frac{1}{2}$.
\end{itemize}
The client runs $A$. If $a=b$, the client gives the server
the reward $\$=1$. If $a\neq b$, $\$=0$.
The expectation value $\langle \$\rangle_b$ of server's reward 
when the server sends $b\in\{0,1\}$ to the client is
\begin{eqnarray*}
\langle \$\rangle_b
&=&1\times{\rm Pr}[a=b]+0\times{\rm Pr}[a\neq b]\\
&=&{\rm Pr}[a=b].
\end{eqnarray*}
Therefore, if $x\in L$,
\begin{eqnarray*}
\langle \$\rangle_{b=1}&=&{\rm Pr}[a=1]>\frac{1}{2},\\
\langle \$\rangle_{b=0}&=&{\rm Pr}[a=0]<\frac{1}{2}.
\end{eqnarray*}
If $x\notin L$,
\begin{eqnarray*}
\langle \$\rangle_{b=0}&=&{\rm Pr}[a=0]>\frac{1}{2},\\
\langle \$\rangle_{b=1}&=&{\rm Pr}[a=1]<\frac{1}{2}.
\end{eqnarray*}
This means that the rational sever wants to send the correct solution
to the client.

Although we use the class PP, the important point here is that the server's
computational ability is enough to be BQP in our protocol.
It is clear that the same proof holds for other classes
in PP, such as AWPP, QCMA, QMA, SBQP, and ${\rm C}_={\rm P}$, etc.

\subsection{Second protocol}
\label{Sec:2ndprotocol}
Let us explain our second protocol,
which is for estimating output probability distributions
of quantum computing.
Consider an $n$-qubit quantum circuit $V$.
Without loss of generality, we can assume that
$V$ consists of only classical gates (such as $X$, CNOT, 
Toffoli, etc.) and Hadamard gates~\cite{HT1,HT2}.
(Generalizations to other gate sets, such as Clifford plus $T$,
are given in Sec.~\ref{App:CliffordplusT}.)
Let
\begin{eqnarray*}
p_z\equiv\langle0^n|V^\dagger(|z\rangle\langle z|\otimes I^{\otimes n-k})
V|0^n\rangle
\end{eqnarray*}
be the probability of obtaining
$z\in\{0,1\}^k$ when the first $k$ qubits of $V|0^n\rangle$
is measured in the computational basis.
We assume that $k=O(\log(n))$.
The client wants to know a 
probability distribution 
$\tilde{p}\equiv\{\tilde{p}_z\}_{z\in\{0,1\}^k}$, which is
close to 
$p\equiv\{p_z\}_{z\in\{0,1\}^k}$ in the sense that
$
|p_z-\tilde{p}_z|\le\frac{1}{poly(n)}
$
for all $z\in\{0,1\}^k$.
Such an estimation $\tilde{p}$ can be obtained
in quantum $poly(n)$ time (see Sec~\ref{App:estimation}), 
but the client who is completely
classical cannot do it by him/herself. The client therefore delegates
the task to the server.
We here provide a protocol where the client can receive 
such an estimation from the rational server.

From $V$, we construct the $(n+1)$-qubit quantum circuit
\begin{eqnarray*}
W_z&\equiv&(I\otimes V^\dagger)
\Big[
\Big(I\otimes|z\rangle\langle z|
+X\otimes(I^{\otimes k}-|z\rangle\langle z|)
\Big)\otimes I^{\otimes n-k}\Big]
(I\otimes V)
\end{eqnarray*}
for each $z\in\{0,1\}^k$.
It is easy to see
\begin{eqnarray}
\langle0^{n+1}|W_z|0^{n+1}\rangle=p_z.
\label{2ndWp}
\end{eqnarray}

\begin{figure}[htbp]
\begin{center}
\includegraphics[width=0.4\textwidth]{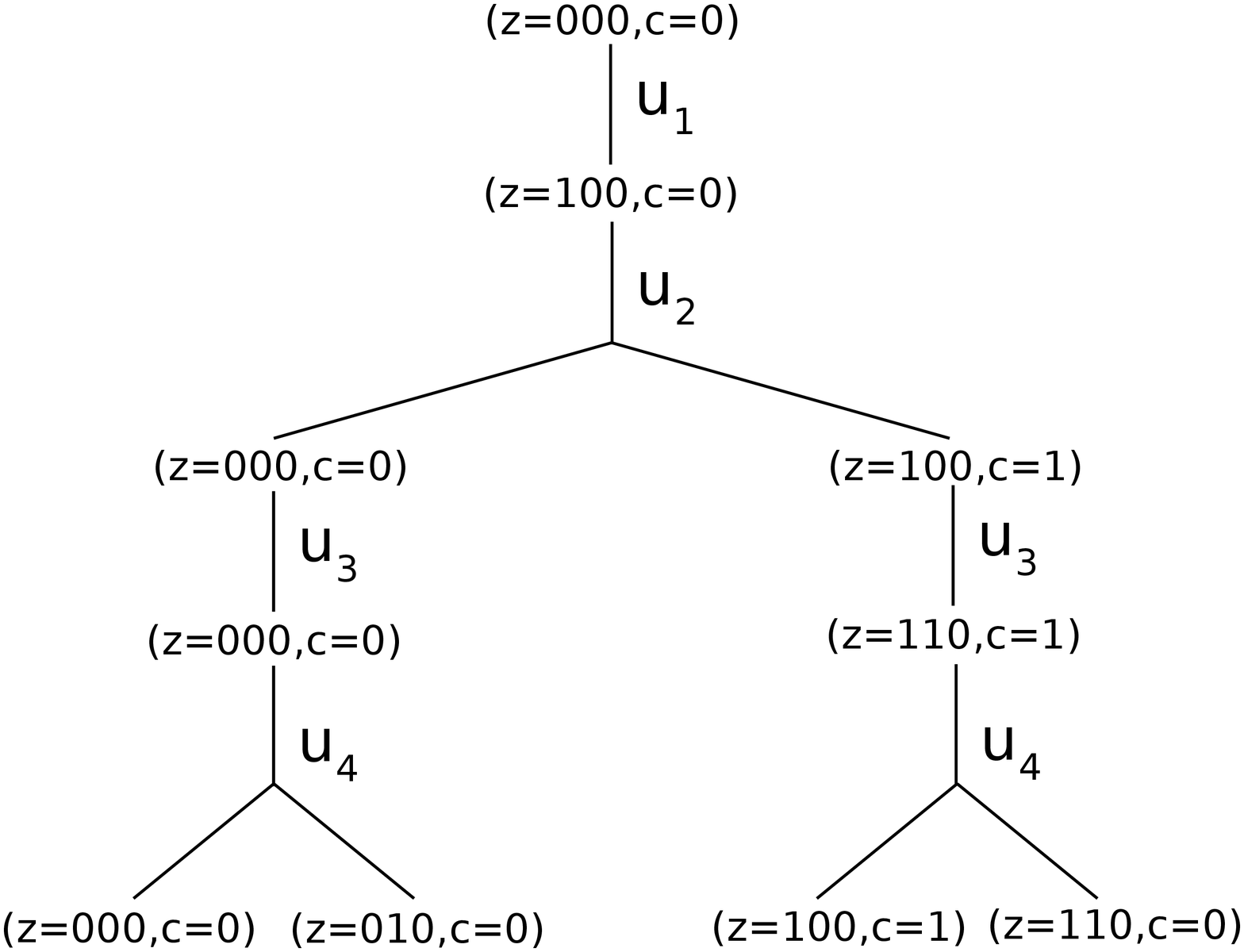}
\end{center}
\caption{
An example of the computational tree.
} 
\label{example}
\end{figure}

We can construct
a classical probabilistic
computing $M_z$ that ``simulates" $W_z$ such that
\begin{eqnarray}
\langle0^{n+1}|W_z|0^{n+1}\rangle=2^h(D_z(1)-D_z(2)),
\label{2ndWD}
\end{eqnarray}
where $h$ is the number of Hadamard gates in $V$,
and $D_z(w)$ 
is the probability that
$M_z$ outputs $w\in\{1,2,3\}$. 
In fact,
Let $t$ be the number of elementary gates
in $W_z$. In other words,
$
W_z=u_t\cdots u_1,
$
where $u_i$ ($i=1,2,...,t$) is a classical gate or the Hadamard gate.
We consider the following $t$-step classical probabilistic
computing $M_z$:
\begin{itemize}
\item[1.]
The state of the register is represented
by the pair $(z,c)$ of an $(n+1)$-bit string 
$z\equiv(z_1,...,z_{n+1})\in\{0,1\}^{n+1}$ and a single 
bit $c\in\{0,1\}$.
The initial state of the register is 
$(z=0^{n+1},c=0)$.
\item[2.]
For $i=1,2,...,t$, do the following:
\begin{itemize}
\item[2-a.]
If $u_i$ is a classical gate, update 
the register as
$(z,c)\to (u_i(z),c)$.
\item[2-b.]
If $u_i$ is the Hadamard gate acting on $j$th qubit,
flip a fair coin. If heads, update the register as
\begin{eqnarray*}
(z,c)\to
(z_1,...,z_{j-1},0,z_{j+1},...,z_{n+1},c).
\end{eqnarray*}
If tails, update the register as
\begin{eqnarray*}
~~~~~~(z,c)\to
(z_1,...,z_{j-1},1,z_{j+1},...,z_{n+1},c\oplus z_j).
\end{eqnarray*}
\end{itemize}
\item[3.]
If the state of the register 
is $(0^{n+1},0)$, output 1.
If the state of the register 
is $(0^{n+1},1)$, output 2.
Otherwise, output 3.
\end{itemize}
An example of the computational tree
for $n=2$, $t=4$, and 
\begin{eqnarray*}
u_1&=&X\otimes I\otimes I,\\
u_2&=&H\otimes I\otimes I,\\
u_3&=&(|0\rangle\langle0|\otimes I+|1\rangle\langle1|\otimes X)\otimes I,\\
u_4&=&I\otimes H\otimes I,
\end{eqnarray*}
is given in Fig.~\ref{example}.

Now it is easy to check Eq.~(\ref{2ndWD}).
(In the example of Fig.~\ref{example}, $D_z(1)=\frac{1}{4}$, $D_z(2)=0$,
and $D_z(3)=\frac{3}{4}$.)
Therefore, from Eqs.~(\ref{2ndWp}) and (\ref{2ndWD}), we obtain
\begin{eqnarray}
p_z=2^h(D_z(1)-D_z(2)).
\label{2ndpD}
\end{eqnarray}

Our protocol runs as follows.
\begin{itemize}
\item[1.]
The server sends 
a classical description of a probability distribution
$p'\equiv\{p_z'\}_{z\in\{0,1\}^k}$ to the client.
If the server is honest, $|p_z'-p_z|\le\frac{1}{poly(n)}$
for all $z$.
\item[2.]
The client chooses $z\in\{0,1\}^k$ uniformly at random.
\item[3.]
The client runs $M_z$ and obtains the output
$w\in\{1,2,3\}$.
\item[4.]
The client pays the reward to the server 
whose amount $\$$ is determined according 
to the following rule:
\begin{itemize}
\item
If $w=1$, then $\$=S(z,p')+2$.
\item
If $w=2$, then $\$=-S(z,p')+2$.
\item
Otherwise, $\$=2$.
\end{itemize}
Here,
\begin{eqnarray*}
S(z,p')\equiv 2p'_z-\sum_{\alpha\in\{0,1\}^k} 
(p'_\alpha)^2-1
\end{eqnarray*}
is called Brier's scoring rule~\cite{Brier}.
\end{itemize}
The expectation value $\langle \$\rangle$
of the prover's reward 
is
\begin{eqnarray*}
\langle \$\rangle
&=&\frac{1}{2^{k+h}}\sum_{z\in\{0,1\}^k}p_zS(z,p')+2,
\end{eqnarray*}
where we have used Eq.~(\ref{2ndpD}).

Note that
\begin{eqnarray*}
\sum_{z\in\{0,1\}^k}p_z S(z,p)
-\sum_{z\in\{0,1\}^k}p_z S(z,p')
=\sum_{z\in\{0,1\}^k}(p_z-p'_z)^2,
\end{eqnarray*}
which means that
$\langle \$\rangle$
is larger if $p'$ is closer to $p$. 
As is explained in Sec.~\ref{App:estimation},
the quantum polynomial-time prover can send $p'$ such that
$|p'_z-p_z|\le\frac{1}{poly(n)}$ for all $z$.
If the server sends another $p'$ such that
$|p'_z-p_z|= const.$
for a certain $z$, on the other hand,
his/her expected profit becomes smaller.
Therefore the rational prover will not do that.

\section{Discussion}
\label{Sec:discussion}
In this paper, we have constructed 
delegated quantum computing protocols
with a classical client and a rational quantum server.
Let us here mention three advantages of our protocols.

First, our protocols are zero-knowledge, which means that
no information other than the solution of the problem itself is leaked 
from the server to the client.

Second, our protocols do not require any extra computational overhead
for the server. For example, 
in the verification protocols of Refs.~\cite{FK,Aharonov}, some extra
trap qubits are needed, and in the verification protocols of
Refs.~\cite{MNS,posthoc,plugin}, the server has to generate the 
Feynman-Kitaev history state
\begin{eqnarray*}
\frac{1}{\sqrt{T+1}}\sum_{t=0}^T(v_t\cdots v_1|0^n\rangle)\otimes|t\rangle,
\end{eqnarray*}
where $V=v_T\cdots v_1$,
which is more complicated than the mere
output state, $V|0^n\rangle$, of the quantum computation. 
On the other hand, in our protocols, what the server has to do is
only the original quantum computing that the client would do
if the client had his/her own quantum computer.

Finally, our protocols neither generate any extra communication overhead
between the server and the client.
In the verification protocols of Refs.~\cite{FK,Aharonov}, 
polynomially many bits have to be exchanged between the server
and the client in order to verify that the server did the correct
measurements on trap qubits.
In the verification protocols of
Refs.~\cite{MNS,posthoc,plugin}, the server has to send the client
a Feynman-Kitaev history state, which consists of polynomially
many qubits.
On the other hand, in our protocols, what the server has to send to the
client is only the solution of the problem that the client
wants to solve.

In our protocols, reward gaps are  
exponentially small. 
It is an open problem whether the constant (or at least polynomial-inverse)
reward gap is possible.
Unfortunately, we can show that as long as we consider a single-round
protocol with the server sending a single bit, it is not possible
unless ${\rm BQP}={\rm BPP}$.
It is shown by using
Theorem 16 of Ref.~\cite{Pavel},
but for readers' convenience, we give a proof here
in our notation.
The expectation value of the server's reward 
when the server sends $b\in\{0,1\}$ to the client is
\begin{eqnarray*}
\langle \$\rangle_b=\sum_wp_w\$(b,w),
\end{eqnarray*}
where the classical probabilistic polynomial-time computing outputs
$w$ with probability $p_w$.
Since $|\$(w,b)|\le const.$ for any $w$ and $b$, the client
can estimate the value of
$\langle \$\rangle_b$ within a $\frac{1}{poly}$ precision
in classical probabilistic polynomial time
by using the standard Chernoff-Hoeffding bound argument.
In fact, let $w_1,w_2,...,w_T$ be the random numbers sampled from 
the probability distribution $\{p_w\}_w$.
The quantity
\begin{eqnarray*}
\eta\equiv \frac{1}{T}\sum_{i=1}^T \$(w_i,b)
\end{eqnarray*}
is an $\epsilon$ precision estimator of 
$\langle \$\rangle_b$ due to the Chernoff-Hoeffding bound:
\begin{eqnarray*}
{\rm Pr}\Big[|\eta-\langle\$\rangle_b|\ge\epsilon]
\le 2\exp\Big[-\frac{T\epsilon^2}{2M^2}\Big],
\end{eqnarray*}
where 
$
M\equiv \max_{w,b}|\$(w,b)|.
$
If $M\le poly$, $T=poly$ is enough to get the $\epsilon=\frac{1}{poly}$
precision.
If $|\langle\$\rangle_{b=1}-\langle\$\rangle_{b=0}|\ge\frac{1}{poly}$,
the client can learn which $b$ gives
larger $\langle \$\rangle_b$ by itself
in classical probabilistic polynomial time,
which means ${\rm BQP}={\rm BPP}$.

One might notice that the above argument does not work if the
server sends the client 
not a single bit $b\in\{0,1\}$ but a polynomial-length bit
string $b\in\{0,1\}^{poly(|x|)}$.
In this case, it is no longer possible
to calculate $\langle \$\rangle_b$ for
all exponentially many $b$ in classical polynomial time.
However, such a generalization does not help,
because, as is shown in Sec.~\ref{App:PtoMA},
the power of such a rational proof system
is in the third level of the
polynomial-time hierarchy.

We also remark effects of errors. For the first protocol, errors in the server
do not cause any problem as long as the bit $b$ is correct.
For the second protocol, again, errors do not cause any problem
as long as the final estimated probabilities are $1/poly$-close to the
true values.

To conclude this paper, let us also mention security of our protocols.
In our first protocol, if the client's result $a$ is leaked to the server
before the server sends $b$ to the client,
the server can cheat. 
Therefore, the client's result $a$ should be hidden from
the server.
On the other hand, a malicious client can cheat the server.
For example, if the server sends $b=1$ to the client,
the malicious client will claim that he/she has generated
$a=0$ thus avoiding the payment.
One way of preventing it would be that the client first commits
$a$ to the server by using the bit commitment protocol.
In this case, however, the security becomes a computational one.

\section{Appendix}

\subsection{Brief summary of Ref.~\cite{AzarMicali}}
\label{App:AM}
Here we briefly summarize some of results in
Ref.~\cite{AzarMicali}.
To understand the essence,
let us consider the following protocol:
\begin{itemize}
\item[1.]
The client samples $w$ from a probability distribution $D$.
\item[2.]
The server sends the client the description of a probability distribution
$D'$.
\item[3.]
The client gives the server the reward $S(D',w)$.
\end{itemize}
Here,
\begin{eqnarray*}
S(D',w)\equiv 2D'(w)-\sum_\alpha(D'(\alpha))^2-1
\end{eqnarray*}
is called Brier's scoring rule~\cite{Brier}.
In the above protocol, server's expected profit is
$
\sum_w D(w)S(D',w).
$
By the straightforward calculation,
\begin{eqnarray*}
\sum_wD(w)S(D,w)-\sum_wD(w)S(D',w)
=\sum_w(D(w)-D'(w))^2.
\end{eqnarray*}
Therefore, server's expected profit is maximum when $D'=D$.
In other words,
if the server wants to maximize the
expected profit, he/she has to send $D'=D$.
The point is that this protocol enables the client, 
who can sample from $D$ but does
not know the description of $D$, to learn the description of $D$
from the rational server.

In Ref.~\cite{AzarMicali}, this idea was used to
construct a single-message rational protocol for
$\#$P problems.
Let 
\begin{eqnarray*}
\phi:\{0,1\}^n\ni x\mapsto \phi(x)\in \{0,1\}
\end{eqnarray*}
be a Boolean function that can be calculated in
classical polynomial time.
The client first samples an $n$-bit string
$x\in\{0,1\}^n$ uniformly at random.
He/She then outputs $\phi(x)$. The probability that
the client outputs 0 is $\frac{\#\phi}{2^n}$,
where $\#\phi$ is the number of $x\in\{0,1\}^n$ such that $\phi(x)=0$.
In other words,
the client can sample from the probability distribution
\begin{eqnarray*}
D:\{0,1\}\ni w\mapsto D(w)\in[0,1]
\end{eqnarray*}
such that $D(0)=\frac{\# \phi}{2^n}$
and $D(1)=1-\frac{\# \phi}{2^n}$.
The ability of sampling from $D$ is not enough for
the BPP client to learn $\#\phi$, since the
estimation of $D(0)$ with an exponential precision is required.
However, if the client uses the above protocol,
the client can learn $\#\phi$, since the rational server
sends the client the description of $D'$ such that
$D'(0)=\frac{\#\phi}{2^n}$ 
and $D'(1)=1-\frac{\#\phi}{2^n}$.

\subsection{Another gate set}
\label{App:CliffordplusT}
\begin{figure}[htbp]
\begin{center}
\includegraphics[width=0.3\textwidth]{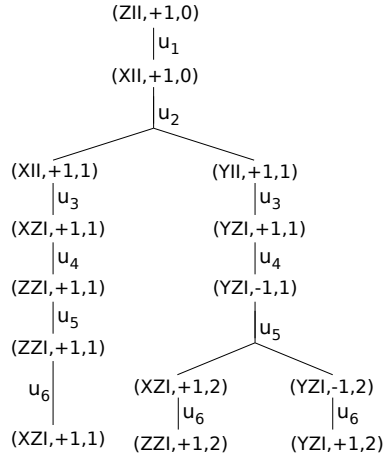}
\end{center}
\caption{An example of the non-deterministic computation. 
For simplicity, the symbol $\otimes$ is omitted,
i.e., $Z\otimes I\otimes I$ is written as $ZII$,
for example.} 
\label{example_T}
\end{figure}

Let us assume that a circuit $V$ consists of only Clifford and 
$T\equiv Z^{\frac{1}{4}}$ gates.
In other words, $V=u_t\cdots u_1$, 
where $u_i$ ($i=1,2,...,t$) is 
$H$, $CZ$, $S=\sqrt{Z}$, or $T$.
Let us consider the following $t$-step
non-deterministic computing:
\begin{itemize}
\item[1.]
The state of the register is represented by $(p,c,k)$,
where $p$ represents the tensor product of $n$ Pauli operators,
$c\in\{+1,-1\}$ represents the sign, 
and $k$ is an integer that counts the number of non-deterministic
transitions experienced.
The initial state of the register is 
$(p=Z\otimes I^{\otimes n-1},c=+1,k=0)$.
\item[2.]
For $i=1,2,...,t$, do the following:
\begin{itemize}
\item[2-a.]
If $u_i$ is a Clifford gate $g$, update the register as
$(p,c,k)\to (g^\dagger pg,c',k)$.
\item[2-b.]
If $u_i$ is $T$ gate acting on $j$th qubit,
and if $j$th Pauli operator of $p$ is $Z$,
do nothing on the register.
\item[2-c.]
If $u_i$ is $T$ gate acting on $j$th qubit,
and if $j$th Pauli operator of $p$ is $X$,
do the following non-deterministic transition:
\begin{eqnarray*}
(p,c,k)\to
\left\{
\begin{array}{ll}
(p_1\otimes...\otimes p_{j-1}\otimes X\otimes p_{j+1}\otimes...\otimes
p_n,c,k+1)\\
(p_1\otimes...\otimes p_{j-1}\otimes Y\otimes p_{j+1}\otimes...\otimes
p_n,c,k+1).
\end{array}
\right.
\end{eqnarray*}
\item[2-d.]
If $u_i$ is $T$ gate acting on $j$th qubit,
and if $j$th Pauli operator of $p$ is $Y$,
do the following non-deterministic transition:
\begin{eqnarray*}
(p,c,k)\to
\left\{
\begin{array}{ll}
(p_1\otimes...\otimes p_{j-1}\otimes X\otimes p_{j+1}\otimes...\otimes
p_n,-c,k+1)\\
(p_1\otimes...\otimes p_{j-1}\otimes Y\otimes p_{j+1}\otimes...\otimes
p_n,c,k+1).
\end{array}
\right.
\end{eqnarray*}
\end{itemize}
\end{itemize}
An example for 
$n=3$, $t=6$,
and
\begin{eqnarray*}
u_1&=&H\otimes I\otimes I,\\
u_2&=&T\otimes I\otimes I,\\
u_3&=&CZ\otimes I,\\
u_4&=&H\otimes I\otimes I,\\
u_5&=&T\otimes I\otimes I,\\
u_6&=&H\otimes I\otimes I,
\end{eqnarray*}
is given in Fig.~\ref{example_T}.

It is easy to check that
\begin{eqnarray*}
\langle0^n|V^\dagger (Z\otimes I^{\otimes n-1})V|0^n\rangle
=\sum_{i:path}\frac{c_i}{\sqrt{2^{k_i}}}f_i,
\end{eqnarray*}
where the summation is taken over all paths,
$(p_i,c_i,k_i)$ is the final state of the register corresponding
to the path $i$,
and
\begin{eqnarray*}
f_i\equiv
\left\{
\begin{array}{ll}
1&\mbox{if $p_i$ consists of only $Z$ and $I$,}\\
0&\mbox{otherwise}.
\end{array}
\right.
\end{eqnarray*}
Hence
\begin{eqnarray*}
p_{acc}&=&
\frac{1}{2}+\frac{1}{2}
\langle0^n|V^\dagger(Z\otimes I^{\otimes n-1})V|0^n\rangle\\
&=&\frac{1}{2}+\frac{1}{2}\sum_{i:path}\frac{c_i}{\sqrt{2^{k_i}}}f_i.
\end{eqnarray*}

\subsection{Estimation}
\label{App:estimation}
We generate $V|0^n\rangle$ and measure the first $k$ qubits
in the computational basis. Output $X=1$ if the result is $z$.
Otherwise, output $X=0$.
We repeat it for $T$ times to correct $X_1,...,X_T\in\{0,1\}$.
If we define
\begin{eqnarray*}
\eta_z\equiv \frac{1}{T}\sum_{i=1}^T X_i,
\end{eqnarray*}
it satisfies
\begin{eqnarray*}
{\rm Pr}[|\eta_z-p_z|\ge\epsilon]
\le 2e^{-2T\epsilon^2}
\end{eqnarray*}
due to the Chernoff-Hoeffding bound.
If we take $\epsilon=\frac{1}{2^kn}$, $T=poly(n)$ is enough to
guarantee that $|\eta_z-p_z|\le \frac{1}{2^kn}$ except for an
exponentially small probability.
We do this procedure for all $z\in\{0,1\}^k$ to obtain
$\{\eta_z\}_{z\in\{0,1\}^k}$.
Except for an exponentially small probability,
$|\eta_z-p_z|\le \frac{1}{2^kn}$ for all $z$.
Let us define 
\begin{eqnarray*}
\tilde{p}_z\equiv\frac{\eta_z}{\sum_{z\in\{0,1\}^k}\eta_z}
\end{eqnarray*}
for each $z$. Then,
\begin{eqnarray*}
\tilde{p}_z\le\frac{p_z+\epsilon}{1-2^k\epsilon}
\le p_z+5\times 2^k\epsilon,
\end{eqnarray*}
and
\begin{eqnarray*}
\tilde{p}_z\ge\frac{p_z-\epsilon}{1+2^k\epsilon}
\ge p_z-5\times 2^k\epsilon.
\end{eqnarray*}
Therefore
\begin{eqnarray*}
|\tilde{p}_z-p_z|\le \frac{1}{poly(n)}
\end{eqnarray*}
for all $z$ except for an exponentially small probability.

\subsection{Longer message}
\label{App:PtoMA}
Let $L$ be a language and $x$ be its instance. 
Assume that $L$ has the following rational proof system:
\begin{itemize}
\item[1.]
The server sends $b\in\{0,1\}^m$ to the client, where $m=poly(|x|)$.
\item[2.]
The client samples a polynomial-length bit string
$w$ from a probability distribution $D$,
and sends the reward $\$(b,w)$ to the server.
\item[3.]
The client calculates a predicate 
$\pi(x,b)\in\{0,1\}$ and accepts/rejects
if $\pi(x,b)=1/0$, where $\pi$ is a polynomial-time computable
Boolean function.
\end{itemize}
The expectation value of the server's reward when he/she sends $b$ to
the client is
\begin{eqnarray*}
\langle\$\rangle_b=\sum_wD(w)\$(b,w).
\end{eqnarray*}
We require that the rational proof system satisfies the following:
\begin{itemize}
\item
When $x\in L$ then
there exists
$b^*\in\{0,1\}^m$ such that
$\pi(x,b^*)=1$, and
$\langle\$\rangle_{b^*}-\langle\$\rangle_b\ge\frac{1}{h}$ for
all $b$ that satisfies $\pi(x,b)=0$,
where $h=poly(|x|)$.
\item
When $x\notin L$ then
there exists
$b^*\in\{0,1\}^m$ such that
$\pi(x,b^*)=0$, and
$\langle\$\rangle_{b^*}-\langle\$\rangle_b\ge\frac{1}{h}$ for
all $b$ that satisfies $\pi(x,b)=1$.
\end{itemize}

We can show that if $\max_{b,w}|\$(b,w)|\le poly(|x|)$,
then $L$ is in 
${\rm NP}^{\rm MA[1]}$, 
which is in ${\rm NP}^{\rm MA}\subseteq\Sigma_3^{\rm P}$~\cite{complexityzoo}.
It means that the above rational proof system will not
contain BQP, because BQP is not believed to be in the third level
of the polynomial-time hierarchy.

In fact, let us consider the following probabilistic polynomial-time
algorithm $M$ on input $(x,b,a)\in\{0,1\}^*\times\{0,1\}^m\times\{0,1\}^m$:
\begin{itemize}
\item[1.]
Calculate $\pi(x,b),\pi(x,a)\in\{0,1\}$.
\item[2.]
By using the Chernoff bound,
calculate $\frac{1}{h^2}-$precision estimates, $\eta_b$ and $\eta_a$,
of $\langle\$\rangle_b$ and $\langle\$\rangle_a$, respectively.
Except for an exponentially small failure probability $e^{-poly(|x|)}$,
$|\eta_b-\langle\$\rangle_b|\le\frac{1}{h^2}$ and 
$|\eta_a-\langle\$\rangle_a|\le\frac{1}{h^2}$.
\item[3.]
If $\pi(x,b)=\pi(x,a)=1$, accept.
If $\pi(x,b)=1$, $\pi(x,a)=0$, and $\eta_b-\eta_a\ge
\frac{1}{h}-\frac{2}{h^2}$, accept.
Otherwise, reject.
\end{itemize}
Then $L$ satisfies the following:
\begin{itemize}
\item
If $x\in L$ then there exists $b$ such that for all $a$,
$M(x,b,a)$ accepts with probability at least $1-2^{-r}$,
where $r=poly(|x|)$.
\item
If $x\notin L$ then for all $b$ there exists $a$ such that 
$M(x,b,a)$ accepts with probability at most $2^{-r}$.
\end{itemize}
Therefore, $L$ is in ${\rm NP}^{\rm MA[1]}$.

\acknowledgements
We thank the anonymous reviewer for pointing out a simpler proof for
the first protocol.
We thank Keiji Matsumoto, Francois Le Gall,
Seiichiro Tani, and Yuki Takeuchi for discussion,
and Pavel Hub{\'a}{\v c}ek for bringing our attention to Ref.~\cite{Pavel}.
TM is supported by JST PRESTO No.JPMJPR176A,
and the Grant-in-Aid for Young Scientists (B) No.JP17K12637 of JSPS. 
HN is supported by the Grant-in-Aid for Scientific Research 
(A) Nos.26247016, 16H01705, (B) No.19H04066,
and (C) No.16K00015 of JSPS.

\end{document}